\documentclass[twocolumn,trackchanges]{aastex7}
\usepackage{bm}
\usepackage{amsmath}
\usepackage{booktabs}
\usepackage{tabularx}
\usepackage{array}

\begin{document}

\title{Exploring Cosmological Constraints of the Void-Lensing Cross-Correlation in the CSST Photometric Survey}

\correspondingauthor{Yan Gong}
\email{Email: gongyan@bao.ac.cn}

\author[orcid=0009-0009-1369-2476]{Qi Xiong}
\affiliation{National Astronomical Observatories, Chinese Academy of Sciences, 20A Datun Road, Beijing 100101, China}
\affiliation{School of Astronomy and Space Sciences, University of Chinese Academy of Sciences(UCAS),\\Yuquan Road NO.19A Beijing 100049, China}
\email{xiongqi@bao.ac.cn}

\author[orcid=0000-0003-0709-0101]{Yan Gong*}
\affiliation{National Astronomical Observatories, Chinese Academy of Sciences, 20A Datun Road, Beijing 100101, China}
\affiliation{School of Astronomy and Space Sciences, University of Chinese Academy of Sciences(UCAS),\\Yuquan Road NO.19A Beijing 100049, China}
\affiliation{Science Center for China Space Station Telescope, National Astronomical Observatories, Chinese Academy of Science, \\20A Datun Road, Beijing 100101, China}
\email{gongyan@bao.ac.cn}

\author{Junhui Yan}
\affiliation{National Astronomical Observatories, Chinese Academy of Sciences, 20A Datun Road, Beijing 100101, China}
\affiliation{School of Astronomy and Space Sciences, University of Chinese Academy of Sciences(UCAS),\\Yuquan Road NO.19A Beijing 100049, China}
\email{yanjh@bao.ac.cn}

\author[0000-0001-8075-0909]{Furen Deng}
\affiliation{National Astronomical Observatories, Chinese Academy of Sciences, 20A Datun Road, Beijing 100101, China}
\affiliation{School of Astronomy and Space Sciences, University of Chinese Academy of Sciences(UCAS),\\Yuquan Road NO.19A Beijing 100049, China}
\email{frdeng@bao.ac.cn}

\author{Hengjie Lin}
\affiliation{National Astronomical Observatories, Chinese Academy of Sciences, 20A Datun Road, Beijing 100101, China}
\email{hjlin@nao.cas.cn}

\author[0000-0001-7283-1100]{Xingchen Zhou}
\affiliation{National Astronomical Observatories, Chinese Academy of Sciences, 20A Datun Road, Beijing 100101, China}
\email{xczhou@nao.cas.cn}

\author{Xuelei Chen}
\affiliation{National Astronomical Observatories, Chinese Academy of Sciences, 20A Datun Road, Beijing 100101, China}
\affiliation{School of Astronomy and Space Sciences, University of Chinese Academy of Sciences(UCAS),\\Yuquan Road NO.19A Beijing 100049, China}
\affiliation{Department of Physics, College of Sciences, Northeastern University, Shenyang 110819, China}
\affiliation{Centre for High Energy Physics, Peking University, Beijing 100871, China}
\email{xuelei@bao.ac.cn}

\author{Qi Guo}
\affiliation{National Astronomical Observatories, Chinese Academy of Sciences, 20A Datun Road, Beijing 100101, China}
\affiliation{School of Astronomy and Space Sciences, University of Chinese Academy of Sciences(UCAS),\\Yuquan Road NO.19A Beijing 100049, China}
\email{guoqi@nao.cas.cn}

\author{Ming Li}
\affiliation{National Astronomical Observatories, Chinese Academy of Sciences, 20A Datun Road, Beijing 100101, China}
\email{mingli@nao.cas.cn}

\author{Yun Liu}
\affiliation{National Astronomical Observatories, Chinese Academy of Sciences, 20A Datun Road, Beijing 100101, China}
\affiliation{School of Astronomy and Space Sciences, University of Chinese Academy of Sciences(UCAS),\\Yuquan Road NO.19A Beijing 100049, China}
\email{liuyun@nao.cas.cn}

\author{Wenxiang Pei}
\affiliation{Shanghai Key Lab for Astrophysics, Shanghai Normal University, Shanghai 200234, China}
\email{wxpei@shnu.edu.cn}

\begin{abstract}

We investigate the cosmological constraints from the void-lensing cross-correlation assuming the $w$CDM model for the Chinese Space Station Survey Telescope (CSST) photometric survey. Using Jiutian simulations, we construct a mock galaxy catalog to $z=3$ covering 100 deg$^2$ which incorporates the instrumental and observational effects of the CSST. We divide the galaxy sample into seven photometric-redshift (photo-$z$) tomographic bins and identify 2D voids within each bin using the Voronoi tessellation and watershed algorithm. We measure the angular cross-power spectrum between the void distribution and the weak lensing signal, and estimate the covariance matrix via jackknife resampling combined with pseudo-$C_{\ell}$ approach to account for the partial sky correction. We employ the Halo Void Dust Model (HVDM) to model the void-matter cross-power spectrum and adopt the Markov Chain Monte Carlo (MCMC) technique to implement the constraints on the cosmological and void parameters. We find that our method can accurately extract the cosmological information, and the constraint accuracies of some cosmological parameters from the void-lensing analysis are comparable or even tighter than the weak lensing only case. This demonstrates that the void-lensing serves as an effective cosmological probe and a valuable complement to galaxy photometric surveys, particularly for the Stage-IV surveys targeting the high-redshift Universe.

\end{abstract}



\section{Introduction} \label{intro}
The accurate measurements of temperature fluctuations of the cosmic microwave background (CMB) \citep{Hinshaw,planck18,Aiola}, together with the discovery of the accelerating expansion of the Universe by Type Ia supernova (SN Ia) observations \citep{Riess,Perlmutter}, have provided important observational evidence for the establishment of the standard cosmological model. This model describes a spatially flat universe composed of roughly $30\%$ matter and $70\%$ dark energy, whose nature remains poorly understood and continues to be one of the main challenges in modern cosmology. The large-scale structure (LSS) of the Universe, arising from the gravitational growth of primordial density fluctuations, provides a powerful set of tools to probe dark matter and dark energy, which can effectively complement the measurements of SN Ia and CMB \citep{DES3x2pt2}. 

Weak gravitational lensing (WL), which refers to slight distortions of distant background galaxies induced by the gravitational potential along the line of sight, has become a powerful probe of the LSS \citep{Bartelmann}. It enables direct mapping of the dark matter distribution and reveals the influence of dark energy on the growth of structure \citep{Kilbinger}.  In particular, measurements of the two-point correlation function or the angular power spectrum of the weak lensing signal, known as cosmic shear, provide stringent constraints on the key cosmological parameters, especially the total matter density parameter $\Omega_m$ and the amplitude of matter fluctuations $\sigma_8$ \citep{KiDs450,KiDs1000,KiDS-Legacy,DESY1,DESY3,DESY3ps}. Moreover, combinations of weak lensing and other tracers of the LSS, such as galaxy clustering (the so-called 3$\times$2pt probes), offer the means of breaking degeneracies between the parameters, mitigating the systematic effects and therefore enhancing cosmological constraints \citep{DES3x2pt1,DES3x2pt2,KiDs3x2pt1,KiDs3x2pt2,KiDs3x2pt3}. 

Recently, cosmic voids, which are large and underdense regions within the cosmic web, have emerged as a promising cosmological probe for studying the LSS \citep{Weygaert,Pan,Pisani,Moresco,Schuster,song3}. Owing to their low-density environment, void formation and evolution are sensitive to the dark energy equation of state and the total neutrino mass \citep{voidDE1,voidDE2,voidDE3,voidDE4,voidneu1,voidneu2}. Their statistics, such as the void shapes, the void density profile and the void size function (VSF), have already been employed to constrain the cosmological parameters \citep{voidshape,VSF1,song1,song2,song3,voidML}. Additionally, several studies have explored the correlation function of cosmic voids and their cross-correlations with other cosmological probes \citep{Granett,void-g1,void-g2,void-a,song4,song5}. 

In particular, it is possible to measure the cross-correlation between void and weak lensing, and exploit this signal to infer cosmological information, and break the parameter degeneracy \citep{void-wl1,void-wl2,void-wl3,void-wl4,void-wl5,void-wl6}. In this work, we investigate the cosmological constraint power of the void-lensing cross-power spectrum in the photometric survey of the Chinese Space Station Survey Telescope \citep[CSST,][]{zhan1,zhan2,gong19,GongFutrue,gong25}.


We construct a mock galaxy catalog covering a sky area of $\sim$100 deg$^2$ based on the Jiutian simulations \citep{JiuTian}, taking into account the instrumental design and survey strategy of the CSST photometric survey. From this mock catalog, we derive the galaxy redshift distribution and divide it into seven photometric-redshift (photo-$z$) tomographic bins. We apply the Voronoi tessellation and watershed algorithm to identify voids in the two-dimensional angular galaxy distribution within each tomographic bin and obtain the void catalog. Then we measure the  angular cross-power spectrum between the void distribution and the weak lensing signal. Covariance estimation is performed using the jackknife method combined with the pseudo-$C_{\ell}$ approach. Finally, we constrain cosmological and systematic parameters using the Markov Chain Monte Carlo (MCMC) method.

This paper is organized as follows. In Section \ref{mock data}, we introduce the construction of the mock galaxy and void catalogs for the CSST photometric survey, and describe the measurement of the void lensing power spectrum. Section \ref{model} presents the theoretical modeling of the angular power spectrum, including the treatment of relevant systematics. The covariance estimation, the model fitting method and the discussion of constraints on the cosmological and systematic parameters are given in Section \ref{constraints}. Our conclusions are summarized in Section \ref{summary}.

\section{Mock Data} \label{mock data}

\subsection{Simulation}
We employ the Jiutian-1G (JT1G) simulation to serve as the basis for constructing the weak lensing convergence map and generating the mock galaxy catalog. JT1G is one of the high-resolution runs in the state-of-the-art Jiutian N-body simulation suite. It is performed with the L-Gadget3 code \citep{Springel05} and evolves $6144^3$ dark matter particles within a cubic simulation box. The box size is $1h^{-1}$Gpc and the particle mass is about 3.72$\times$10$^8$ $h^{-1}$ $M_{\odot}$. We adopt the cosmological parameters from $\it Planck$2018 \citep{planck18}, with $\Omega_{\text{m}} = 0.3111$, $\Omega_{\text{b}} = 0.0490$, $\Omega_\Lambda = 0.6899$,  $\sigma_8 = 0.8102$, $n_{\text{s}} = 0.9665$, and $h = 0.6766$. The JT1G simulation outputs 128 snapshots from redshift 127 to 0 with an average time gap of $\sim 100$ Myr, which is suitable for weak lensing studies. The dark matter halos and subhalos are identified using the friend-of-friends (FOF) and SUBFIND algorithm \citep{Springel05}. Additionally, the subhalos are linked with their unique descendants to establish the merger trees. Further details of JT1G can be found in \cite{JiuTian}.

Considering the effects of structure evolution, we construct a partial-sky light cone with a set of discrete slices based on the outputting snapshots at different redshifts. To dilute the structure repetitions caused by box replication, a specific viewing angle is chosen as the line-of-sight (LOS) direction, producing a light cone that extends to $z = 3$ and covers a sky area of $100$ deg$^2$, which is sufficient for the purposes of this study. The light-cone construction for both dark matter particles and halos (or galaxies) follows the same strategy, ensuring that different tracers capture identical cosmological information \citep[see e.g.][]{xiong}. 

\subsection{Galaxy Mock Catalog}
\begin{figure}
    \centering
    \includegraphics[width=1\linewidth]{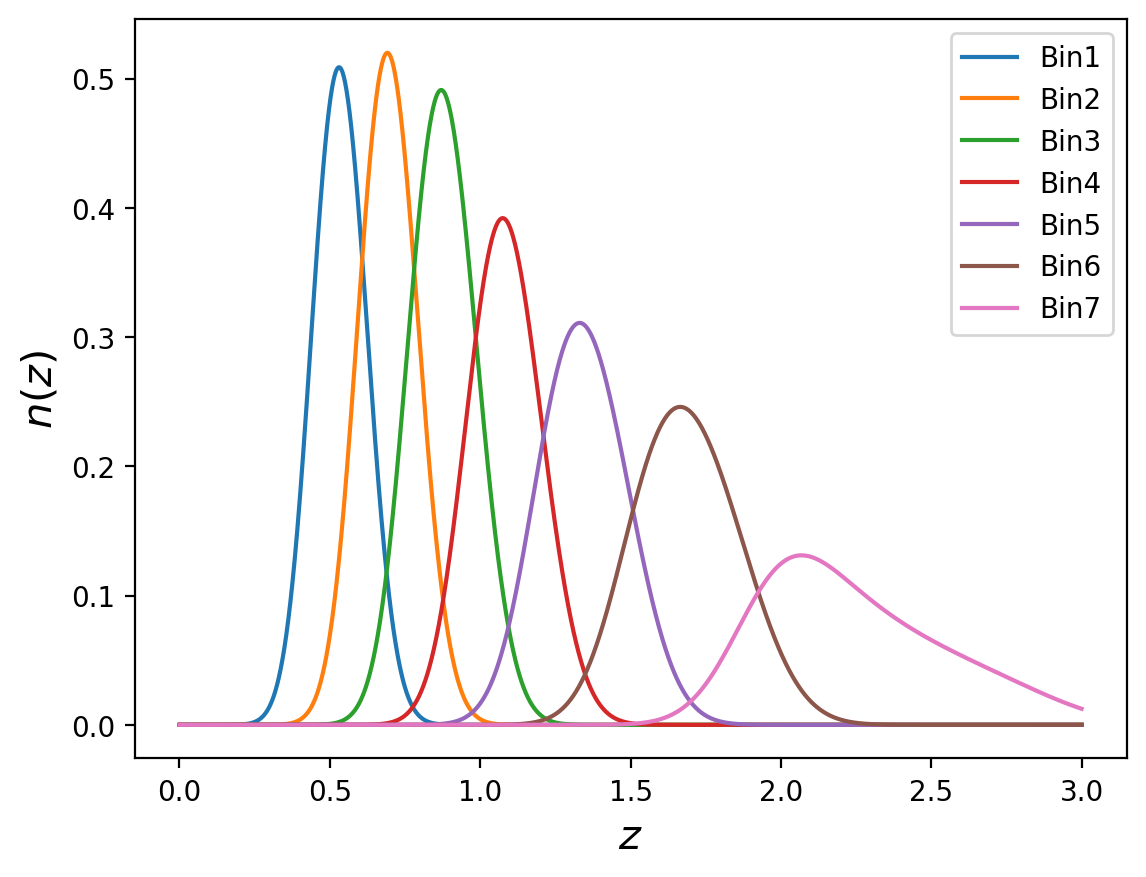}
    \caption{The mock galaxy redshift distributions in the CSST photometric survey. The solid curves denote the redshift distributions of the seven photo-$z$ bins, which are obtained by stacking samples drawn from the redshift PDF of each individual galaxy.}
    \label{fig:red_dis}
\end{figure}

We populate the simulations with galaxies utilizing an improved Semi-Analytic Model \citep{henriques2015galaxy,pei}, which provides a variety of observational properties including galaxy spectral-energy distribution (SED), emission lines, and apparent magnitudes. Galaxy samples are then selected according to the apparent magnitude limits of the CSST photometric survey, which can reach $i \sim 26$ AB mag for $5\sigma$ point source detection \citep{gong19,gong25}. To estimate photometric redshifts, we assume that the observed redshift of each galaxy follows a Gaussian probability distribution function (PDF), $z_{\text{obs}} \sim N(z_{\text{true}}, \sigma_{z})$, where $z_{\text{true}}$ is the true redshift and $\sigma_{z}$ represents the redshift uncertainty characterized as $\sigma_{z} = \sigma_{z_0}(1+z)$ with $\sigma_{z_0} = 0.05$ \citep{gong19}. The resulting photo-$z$ distribution from the mock catalog can be obtained by stacking samples drawn from the redshift PDFs of individual galaxies. In Figure \ref{fig:red_dis}, we show the redshift distributions of the seven photo-$z$ bins in the CSST photometric survey.

We mimic weak lensing distortions, including convergence $\kappa$ and shear $\gamma = \gamma_1 + i\gamma_2$, using the multi-lens-plane algorithm implemented in LensTools\footnote{\url{https://github.com/apetri/LensTools}} \citep{multi-lens,lenstools}. Dark matter particles in the light cone are divided into discrete slices, and light rays are traced from the observer to the source planes to compute deflections and generate synthetic lensing maps. Note that since the resulting maps reside on an observer grid, we shift galaxies to their apparent positions via lensing deflections, and then assign lensing quantities to each galaxy based on its redshift and apparent position \citep{xiong}. 

\setlength{\tabcolsep}{9pt}
\begin{deluxetable}{ccccc}\label{tab:catalog}
\tablecaption{The CSST galaxy and void samples used in this work. The photo-$z$ tomographic bins, number of galaxies and voids in each bin ($N_{\text{g}}$ and $N_{\text{v}}$), and the average surface number density of galaxies and voids in arcmin$^{-2}$ ($\bar{n}_{g}$ and $\bar{n}_{\text{v}}$) have been listed.}
\tablehead{
\colhead{redshift bin} & 
\colhead{$N_{\text{g}}$} & \colhead{$\bar{n}_{\text{g}}$} & \colhead{$N_{\text{v}}$} & \colhead{$\bar{n}_{\text{v}}$} 
}
\startdata
$0.45 < z \leq 0.61$ & 1090894 & 3.03 & 35732 & 0.099 \\
$0.61 < z \leq 0.78$ & 1217180 & 3.38 & 40434 & 0.112 \\
$0.78 < z \leq 0.98$ & 1298654 & 3.61 & 43499 & 0.121 \\
$0.98 < z \leq 1.20$ & 1137921 & 3.16 & 76719 & 0.213 \\
$1.20 < z \leq 1.50$ & 1116709 & 3.10 & 25961 & 0.072 \\
$1.50 < z \leq 1.90$ & 1054731 & 2.93 & 20488 & 0.057 \\
$1.90 < z \leq 3.00$ & 904882  & 2.51 & 12613 & 0.035 \\
\enddata
\end{deluxetable}

\begin{figure*}
    \centering
    \includegraphics[width=0.95\linewidth]{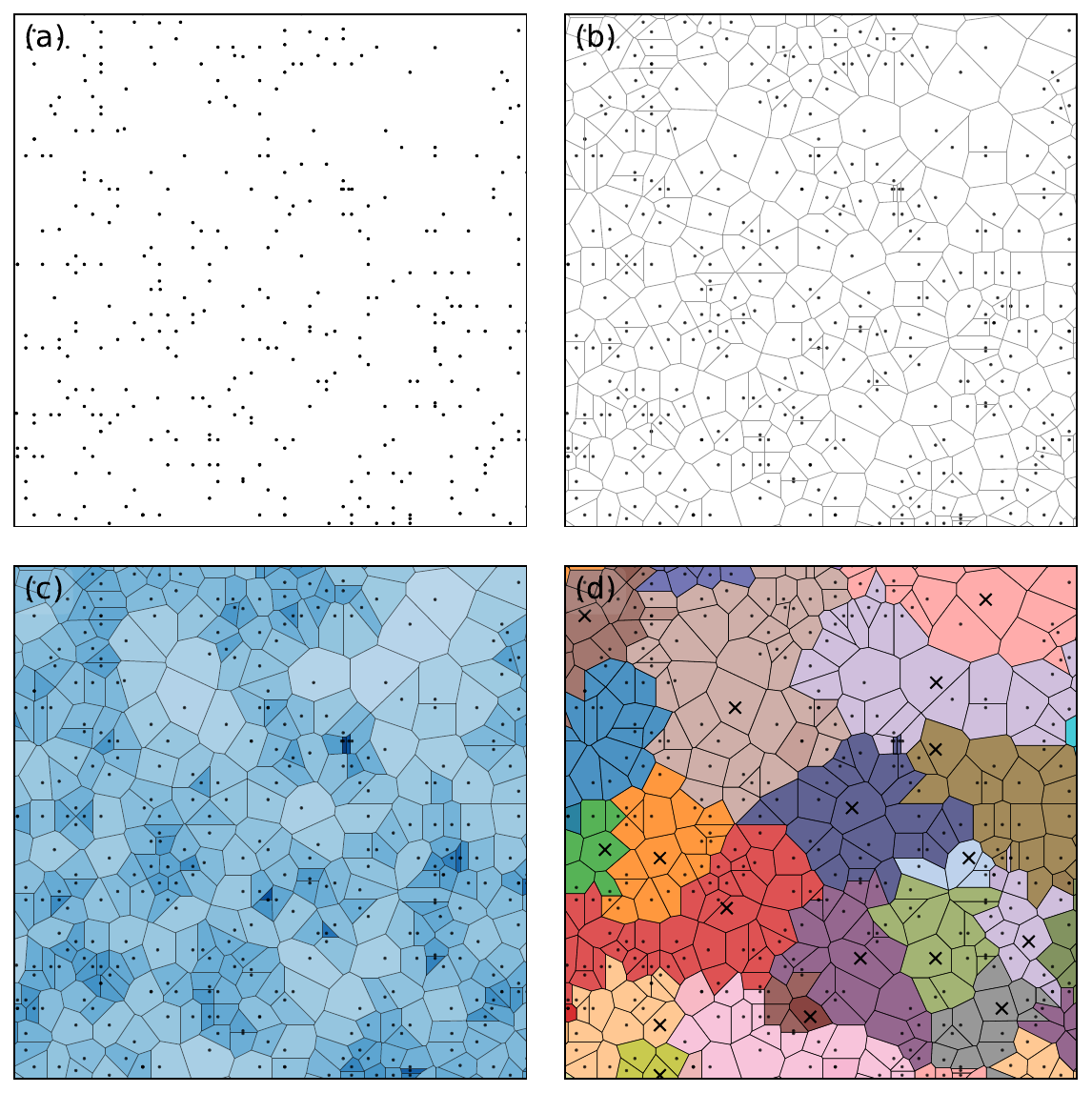}
    \caption{Schematic illustration of the procedure for identifying 2D voids from the galaxy catalog. (a) A slice of galaxies from a small region of the mock catalog, where black dots indicate galaxy positions. (b) The 2D Voronoi tessellation constructed from the galaxy distribution, with each galaxy assigned to a Voronoi cell. (c) Voronoi cells colored by the local density. (d) Identified zones (2D void candidates without pruning), where crosses mark the cores (local density minima) of the zones and colors distinguish different zones.}
    \label{fig:void_id}
\end{figure*}

\subsection{Void Mock Catalog} \label{void cat}
We adopt the Voronoi tessellation and
watershed algorithm to identify voids in the 2D angular distribution of galaxies within projected slices, which have been widely employed in 3D void identification \citep{3Dvoid1,3Dvoid2,3Dvoid3}. To ensure that most galaxies will be assigned to the correct slice, the line-of-sight width of each slice should be sufficiently large, at least about twice the photo-$z$ uncertainty. Specifically, we divide the galaxy sample into seven photo-$z$ tomographic bins, each containing similar numbers of galaxies. The bin edges are defined by the endpoints
$\{z_{e}\} = \{0.45, 0.61, 0.78, 0.98, 1.20, 1.50, 1.90, 3.0\}$, as presented in Table~\ref{tab:catalog}.
This binning strategy aims to extract more information and mitigates the overlap of 2D voids across slices. Here we focus on the redshift range $z=0.45 - 3.0$ for our analysis, since the LSS at lower redshifts is well-measured by spectroscopic surveys and galaxy density decreases rapidly at higher redshifts in the CSST photometric survey \citep{gong19,song5,xiong}.

The algorithm begins by producing a Voronoi tessellation of the galaxy sample, which partitions space into cells around each galaxy. Each cell includes a single galaxy and is defined as the region of space closer to that galaxy than to any other in the sample. The density within each cell is then estimated as $\rho_{\text{cell}} = 1/S_{\text{cell}}$, where $S_{\text{cell}}$ denotes the cell area determined by Voronoi tessellation. The Voronoi tessellation also provides a natural set of neighbors for each galaxy, which can be used to identify local density minima. Zones are then built from each density minimum in the distribution of cells using a watershed algorithm, where each cell is linked to its least dense neighbor and merged accordingly, repeating the process until it arrives at a minimum. Finally, voids are formed by identifying low-density boundaries (saddle points) between adjacent zones and merging them into larger unions of separated regions. In Figure~\ref{fig:void_id}, we illustrate the procedure for identifying 2D voids from the galaxy catalog.

\begin{figure*}
    \centering
    \includegraphics[width=1\linewidth]{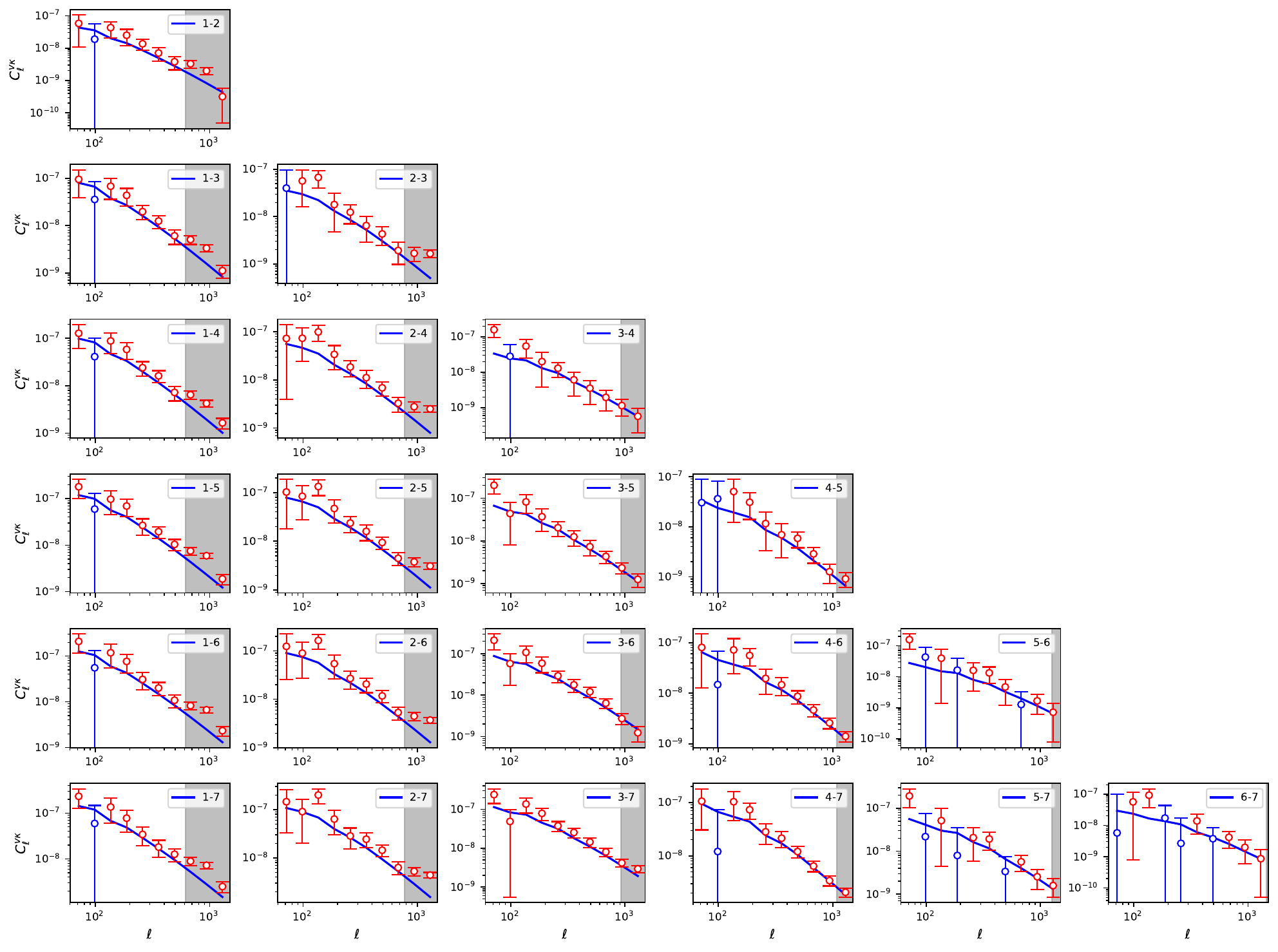}
    \caption{The mock CSST void-lensing cross-power spectra for the seven tomographic bins in 100 deg$^2$. The blue solid curves show the results of the best-fitting theoretical model, and the data points are the mock data. The blue data points denote the signal-to-noise ratio (SNR) $<1$,  which are discarded in the constraint process. The gray regions show the small scales that are excluded with $k_{\text{max}} = 0.3\, $Mpc$^{-1}h$. We also discard the cross-power spectra with low amplitudes, and only consider the void-lensing power spectrum for the case $i < j$ in the analysis, where $i$ and $j$ denote the tomographic bins of the void and weak lensing samples, respectively}
    \label{fig:vk}
\end{figure*}

In 3D void identification, various combinations of zones may be defined as voids, and there exist different treatments for combining zones, referred to as pruning methods \citep{DESIVAST}. To determine which zones should be combined into definitive voids in our 2D void identification, we apply a pruning procedure similar to the VIDE method \citep{3Dvoid2}. Specifically, we combine adjacent zones into a void if density of the boundaries between them is less than 0.4 times the average density, and remove voids with central-density greater than 0.4 times the average density. 
The angular radius of a void $\theta_{\text{v}}$ is then derived from an effective circle whose area $S_{\text{v}}$ equals the total area of all cells comprising the void, which is given by
\begin{equation}
    S_{\text{v}} = \sum_{i}S^{i}_{\text{cell}} =\pi \theta^{2}_{\text{v}},
\end{equation}
where $S^{i}_{\text{cell}}$ denotes the area of the cell $i$. The void position is characterized by its area-weighted center $\mathbf{X}_{\text{v}}$, which is calculated from its constituent cells as 
\begin{equation}
    \mathbf{X}_{\text{v}} = \frac{1}{S_{\text{v}}}\sum_{i}\mathbf{x}_{i}S^{i}_{\text{cell}},
\end{equation}
where $\mathbf{x}_{i}$ is the coordinate of the galaxy within the cell $i$.

\subsection{Void-Lensing Power Spectrum Measurement}

Having obtained the mock catalogs, we can measure the angular cross-power spectrum between the void distribution and the weak lensing signal. We first construct the tomographic shear maps by projecting the shear values of each galaxy onto a two-dimensional Cartesian grid using the formula
\begin{equation}
    \bm{\hat{\gamma}_{p}}(\bm{\theta}) = \frac{\sum_{i\in p}w_{i}\bm{\hat{\gamma}_{i}}(\bm{\theta})}{\sum_{i\in p}w_{i}},
\end{equation}
where the index $i$ runs over galaxies in the catalog, $p$ denotes the map pixels, $\bm{\hat{\gamma}_{i}} = (\hat{\gamma}_{1,i},\hat{\gamma}_{2,i})$ is the shear of the $i$-th galaxy, $\bm{\theta}$ represents the angular coordinates on the sky, and $w_{i}$ is the associated weight. For simplicity, we set $w_{i} = 1$ for all galaxies. The shear field is then converted into a lensing convergence field via the Kaiser-Squires inversion \citep{KS}, which is expressed as
\begin{equation}
    \tilde{\kappa}(\bm{\ell}) = \left (\frac{\ell_{1}^{2} - \ell_{2}^{2}}{\ell_{1}^{2}+\ell_{2}^{2}} \right ) \tilde{\gamma}_{1}(\bm{\ell}) + 2 \left ( \frac{\ell_1 \ell_2}{\ell_{1}^{2}+\ell_{2}^{2}} \right ) \tilde{\gamma}_{2}(\bm{\ell}),
\end{equation}
where $\bm{\ell} = (\ell_1,\ell_2)$ denotes the multipoles in Fourier space, and $\tilde{\kappa}$ and $\bm{\tilde{\gamma}} = (\tilde{\gamma_1},\tilde{\gamma_2})$ are the Fourier transforms of the convergence $\kappa$ and shear $\bm{\gamma}$ fields, respectively. The real space convergence map is then obtained by performing the inverse fast Fourier transform of $\tilde{\kappa}$.


On the other hand, the tomographic void overdensity map can be obtained by
\begin{equation}
    \delta_{p} = \frac{N_{p}\sum_{p^{\prime}}A_{p^{\prime}}}{A_{p}\sum_{p^{\prime}}N_{p^{\prime}}} -1,
\end{equation}
where $N_{p} = \sum_{i\in p}\text{v}_{i}$ is the number of voids at a given pixel $p$, $A_{p}$ is the effective fraction of the area at pixel $p$.

We derive the mock angular cross-power spectrum from the void density maps and weak lensing convergence maps at $\ell < 1500$. The results are shown in Figure \ref{fig:vk}, along with the theoretical prediction calculated using the $w$CDM best-fitting parameters. The covariance matrix is estimated using the jackknife method combined with the pseudo-$C_{\ell}$ approach \citep{Peebles,pseudo1,pseudo2}, as implemented in NAMASTER code\footnote{\url{https://github.com/LSSTDESC/NaM ster}} \citep{pseudo3}. The details of the covariance estimation are provided in Section \ref{sec:cov}. 

\section{Theoretical Modeling} \label{model}
Under the flat sky assumption and the Limber approximation \citep{Limber}, the angular power spectrum $C_{AB}^{ij}(\ell)$ between the $i$-th tomographic bin of probe $A$ and $j$-th tomographic bin of probe $B$ can be expressed as
\begin{equation}
    C_{AB}^{ij}(\ell) = \int d\chi \frac{q_{A}^{i}(\chi)q_{B}^{j}(\chi)}{\chi^2}P_{AB}\left( \frac{\ell+1/2}{\chi},\chi \right),
\end{equation}
where $A$ and $B$ denote two different types of tracers, $\chi$ is the comoving radial distance, $q^{i}(\chi)$ is the weighting function of the $i$-th tomographic bin, and $P_{AB}$ denotes the probe-dependent power spectrum.

For the case of voids and weak lensing, $P_{AB}$ corresponds to the void-matter cross-power spectrum $P_{\text{v}m}$, which is modeled using the Halo Void Dust Model (HVDM) \citep{HVDM}. The fundamental assumption of HVDM is that all matter in the Universe belongs to dark matter halos, cosmic voids, and `dust', which refers to linear matter field between halos and voids. Accordingly, the matter density field can be decomposed as
\begin{equation}
\begin{aligned}
    \rho(\mathbf{x}) = \sum_{i}^{\text{halos}} \rho_h(\mathbf{x} - \mathbf{x}_i \mid M_i) 
&+ \sum_{j}^{\text{voids}} \rho_{\text{v}}(\mathbf{x} - \mathbf{x}_j \mid M_j) \\
&+ \rho_d(\mathbf{x}),
\end{aligned}
\end{equation}
where $\rho_h(\mathbf{x} - \mathbf{x}_i \mid M_i)$ is the density profile of a halo with mass $M_i$ centered at position $\mathbf{x_i}$, $\rho_{\text{v}}(\mathbf{x} - \mathbf{x}_j \mid M_j)$ is the density profile of a void with mass $M_j$ centered at position $\mathbf{x}_j$, and $\rho_d(\mathbf{x})$ is the dust density.

In Fourier space, the void-matter power spectrum can be decomposed into contributions from 1-Void, 2-Void, Halo-Void and Void-Dust terms,  which is expressed as
\begin{equation}
\begin{aligned}
    P_{\text{v}m}(k|M) &=  P_{\text{v}m}^{1V}(k|M)  + P_{\text{v}m}^{2V}(k|M) \\ &+ P_{\text{v}m}^{HV}(k|M) + P_{\text{v}m}^{VD}(k|M).
\end{aligned}
\end{equation}
If we consider large scales, we can simplify the 2-Void, Halo-Void and Void-Dust terms and obtain \citep{HVDM}
\begin{equation}
    P_{\text{v}m}(k|R) = P_{\text{v}m}^{1V}(k|R) + b_{\text{v}}(R)P_{mm}^{L}(k),
\end{equation}
where $b_{\text{v}}$ denotes the void bias, $P_{mm}^{L}$ is the linear matter power spectrum computed using {\tt CAMB} \citep{camb}, and $P_{\text{v}m}^{1V}$ represents the 1-Void term present in the HVDM, which is given by
\begin{equation}
    P_{\text{vm}}^{1V} = \frac{M}{\bar{\rho}_m} \, u_{\text{v}}(k|M),
\end{equation}
where $u_{\text{v}}$ is the Fourier transform of the void profile 
\begin{equation}
    u_{\text{v}}(k|M) = \int_{0}^{r_{\text{v}}}\frac{4\pi r^2}{M}\frac{\text{sin}kr}{kr}\rho_{\text{v}}(r|M)dr.
\end{equation}
Here we adopt the Hamaus-Sutter-Wandelt (HSW) profile \citep{HSW}, and it is given by
\begin{equation}
    \frac{\rho_{\text{v}}(r)}{\bar{\rho}_m} -1 =\delta_{c}\frac{1-(r/r_{s})^{\alpha}}{1+(r/r_{\text{v}})^{\beta}},
\end{equation}
where $\delta_{c}$ is the central density contrast, $r_{s}\equiv\gamma r_{\text{v}}$ is the scale radius at which $\rho_{\text{v}}=\bar{\rho}_m$, and $\alpha$ and $\beta$ determine the inner and outer slopes of the void compensation wall, respectively. These four parameters ($\delta_c, \alpha, \beta, \gamma$) are treated as free parameters in our fitting procedure. For simplicity, we use the mean void radius $R_{\text{v}}^{\text{mean}}=\theta_{\text{v}}^{\text{mean}}D_{A}$ in our calculation, where $D_{A}$ is the comoving angular diameter distance of each tomographic bin.

The weighting functions of void and weak lensing are given by 
\begin{equation} \label{equ:v}
    q^{i}_{\text{v}}(\chi)=n^{i}_{\text{v}}\left(z(\chi)\right)\,\frac{dz}{d\chi},
\end{equation}
\begin{equation} \label{equ:k}
    q_{\kappa}^{i}(\chi) = \frac{3H_{0}^{2}\Omega_{m}}{2c^2}\frac{\chi}{a(\chi)}\int_{\chi}^{\infty} d\chi^{\prime}n^{i}_{g}(z(\chi^{\prime}))\frac{dz}{d\chi^{\prime}}\frac{\chi^{\prime}-\chi}{\chi^{\prime}},
\end{equation}
where $n^{i}_{\text{v}}$ is the normalized void redshift distribution in the $i$-th tomographic bin, $H_0$ is the Hubble constant, $\Omega_m$ is the matter density, $c$ is the speed of light, $a(\chi)$ is the scale factor corresponding to the comoving distance $\chi$, and $n^{i}_{g}$ is the normalized redshift distribution of the source galaxies of the tomographic bin $i$.

When modeling the redshift distribution $n_{g}^{i}(z)$, we account for possible uncertainties by introducing shift parameter $\Delta z^{i}$ and stretch parameter $\sigma_{z}^{i}$ as free parameters \citep{redun1,redun2}. The modified redshift distribution is then given by
\begin{equation}
n_{g}^{i}(z) \rightarrow \frac{n_{g}^{i}}{\sigma_{z}^{i}} \left( \frac{z - \langle z^{i} \rangle - \Delta z^{i}}{\sigma_{z}^{i}} + \langle z^{i} \rangle \right),
\end{equation}
where $\left< z^i \right> =  \int zn_{g}^{i}(z)dz / \int n_{g}^{i}(z)dz$ denotes the mean redshift of the $i$-th tomographic bin.

To account for uncertainties in shear estimation, we introduce multiplicative factors to the weak lensing weighting function as 
\begin{equation}
    q^{i}_{\kappa}(\chi) \rightarrow (1+m^{i})q^{i}_{\kappa}(\chi),
\end{equation}
where $m^{i}$ represents the shear calibration bias for the $i$-th source bin. Furthermore, since the measured cross-power spectra are subject to both statistical and systematic uncertainties, we model these uncertainties by incorporating a noise term, leading to
\begin{equation}
    \tilde{C}^{ij}_{\text{v}\kappa}(\ell) = C^{ij}_{\text{v}\kappa}(\ell) + N^{ij}_{\text{v}\kappa},
\end{equation}
where $N^{ij}_{\text{v}\kappa}$ denotes the noise contribution, which we treat as free parameters in the constraint process. For simplicity, we neglect the contribution from intrinsic alignments in the void-lensing cross-correlation \citep{void-wl5}.

Equations~ (\ref{equ:v}) and (\ref{equ:k}) indicate that the lensing kernel of a high-redshift bin has a large overlap with the void weighting function of the low-redshift bin. Moreover, the weak lensing signal is the accumulative effect along the LOS, and we expect a strong correlation between the background cosmic shear and the foreground void clustering. Therefore, we only consider the void-lensing power spectra for the case $i<j$ to perform the cosmological analysis, as shown in Figure~\ref{fig:vk}, where $i$ and $j$ denote the tomographic bins of the void and weak lensing samples, respectively.

Since the void model is valid on large scales, we apply a scale cut procedure in our void lensing analysis analogous to that used in galaxy-galaxy lensing to mitigate modeling uncertainties at small scales. Specifically, we only consider the multipoles that satisfy $\ell \leq k_{\text{max}}\chi(\left<z^{i}\right>)$ and adopt $k_{\text{max}} = 0.3\, $Mpc$^{-1}h$.

\section{Constraint and Results} \label{constraints}
\subsection{Covariance Estimation} \label{sec:cov}
The covariance matrix and more specifically its inverse, i.e. the precision matrix, is the key quantity that determines the uncertainties on cosmological parameters. In general, it can be estimated through three approaches: estimation from numerical simulations, analytical modeling/computation, and direct estimation from the data \citep{cov1}. 
Here we adopt the third option and estimate the covariance using jackknife resampling.

To compute the jackknife covariance, we first construct a set of jackknife samples by systematically excluding one jackknife segment at a time and using the remaining data to measure the angular power spectrum. However, this procedure inevitably introduces a systematic bias due to the altered survey footprint. This is because the `excluding one segment' operation is equivalent to applying a mask to the map, which leads to mode coupling in the power spectrum. To account for this effect, we estimate the angular power spectrum of each jackknife realization using the so-called pseudo-$C_{\ell}$ method \citep{Peebles,pseudo1,pseudo2}, which is implemented in NAMASTER code \citep{pseudo3}. 

\begin{figure}
    \centering
    \includegraphics[width=1\linewidth]{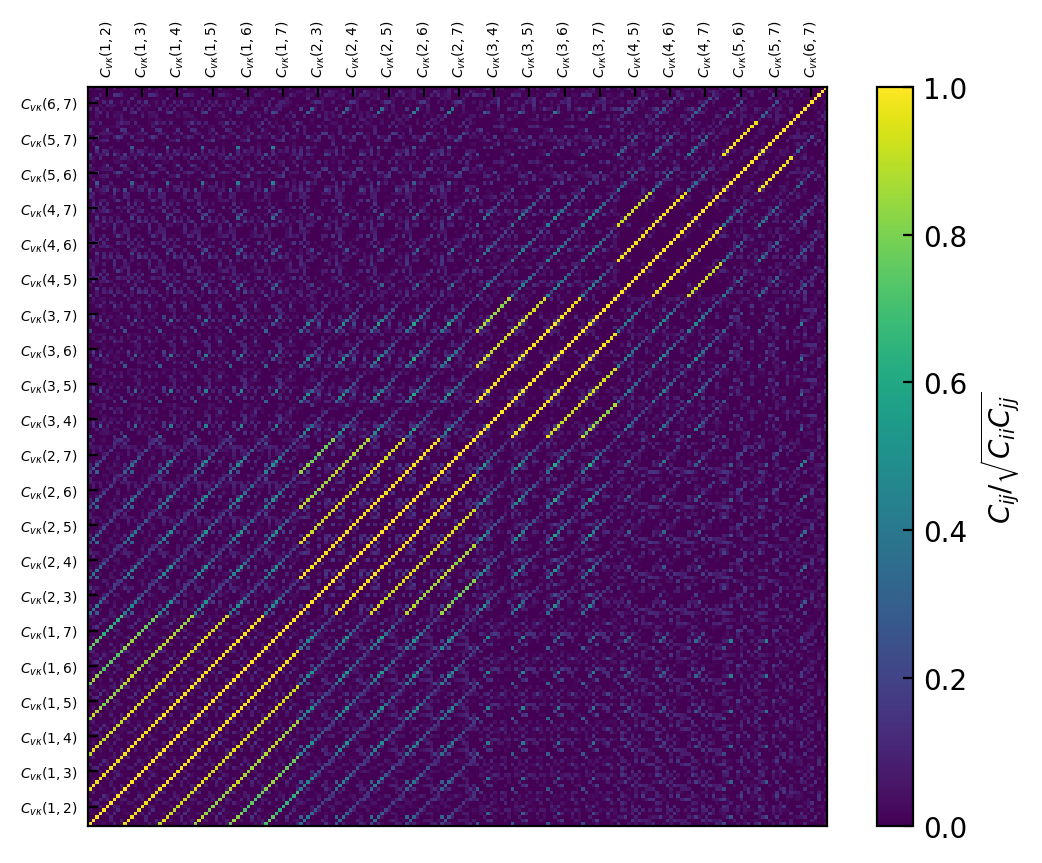}
    \caption{The normalized full covariance matrix, shown in terms of the correlation coefficient $C_{ij}/\sqrt{C_{ii}C_{jj}}$, where $C_{ij}$ are the elements of the covariance matrix, estimated from jackknife realizations. The $C_{\text{v}\kappa}$ values in the data vector are ordered according to their redshift bin combinations $(m, n)$.}
    \label{fig:cov}
\end{figure}

The estimator of the covariance matrix is then given by
\begin{equation}\label{equa:cov}
\begin{aligned}
    \text{Cov}_{\text{JK}}^{ijmn}(\ell_{1},\ell_{2}) &= \frac{N_{\text{JK}}-1}{N_{\text{JK}}} \sum_{k=1}^{N_{\text{JK}}}(C_{k}^{ij} (\ell_{1})-\bar{C}^{ij}(\ell_{1}))^{T}\\
    &\times(C_{k}^{mn}(\ell_{2})-\bar{C}^{mn}(\ell_{2})) ,
\end{aligned}
\end{equation}
where superscript $i$, $j$, $m$, $n$ label the tomographic bins and in combination represent the cross-spectrum between two tomographic bins, $N_{\text{JK}}$ is the number of jackknife samples. To ensure the covariance matrix is non-singular, $N_{\text{JK}}$ must be greater than the length of the data vector \citep{cov2,cov3}. Here $C_{k}^{ij}(\ell)$ denotes the angular power spectrum measured for the $k$-th jackknife realization, and $\bar{C}^{ij}(\ell)$ is the mean spectrum over all jackknife samples, which is given by
\begin{equation}
    \bar{C}^{ij}(\ell) = \frac{1}{N_{\text{JK}}}\sum_{k=1}^{N_{\text{JK}}}C_{k}^{ij}(\ell).
\end{equation}
In Figure~\ref{fig:cov}, we show the normalized covariance matrix estimated from the jackknife realizations, where the block structure reflects correlations among different tomographic bin combinations and multipole ranges.

From Equation~(\ref{equa:cov}), we can obtain the precision matrix estimator \citep{Hartlap}
\begin{equation}\label{cov-1}
    \text{Cov}^{-1} = \frac{N_{\text{JK}}-d-2}{N_{\text{JK}}-1}\text{Cov}_{\text{JK}}^{-1} ,
\end{equation}
where the prefactor corrects for the multiplicative bias resulting from the inversion of the covariance matrix estimator, and $d$ is the length of the data vector.

\setlength{\tabcolsep}{10pt}
\begin{deluxetable}{ccc}\label{tab:fit}
\tablecaption{The fiducial values and priors of the free parameters considered in the constraint process. Uniform priors are described by $\mathcal{U}(x,y)$, with $x$ and $y$ denoting the prior range. The Gaussian priors are represented by $\mathcal{N}$$(\sigma,\mu)$, and $\sigma$ and $\mu$ are the mean and standard deviation, respectively.}
\tablehead{
\colhead{Parameter} & \colhead{Fiducial Value} & \colhead{Prior}
}
\startdata
& \textbf{Cosmology} \\
$h$ & 0.6766 & $\mathcal{U}(0.4,1.0)$ \\
$\Omega_{m}$ & 0.3111 & $\mathcal{U}(0.1, 0.6)$  \\
$10^{9}A_s$ & $2.03$ & $\mathcal{U}(0.5, 5.0)$  \\
$n_{s}$ & 0.9665 & $\mathcal{U}(0.5, 1.5)$ \\
$w$ & -1 & $\mathcal{U}$(-1.8, 0.2) \\
\hline
& \textbf{Void porfile} \\
$\alpha^{i}$ & - &  $\mathcal{U}$(0, 10) \\
$\beta^{i}$ & - &  $\mathcal{U}$(0, 20) \\
$\gamma^{i}$ & - &  $\mathcal{U}$(0, 5) \\
$\delta_{\text{cen}}^{i}$ & - & $\mathcal{U}$(-1, -0.5) \\
\hline
& \textbf{Void bias} \\
$b_{\text{v}}^{i}$ & - & $\mathcal{U}$(0, 5) \\
\hline
& \textbf{Photo-$z$ shift} \\
$\Delta z^{i}$ & (0, 0, 0, 0, 0, 0) & $\mathcal{U}$(-0.1, 0.1) \\
\hline
& \textbf{Photo-$z$ stretch} \\
$\sigma_{z}^{i}$ & (1, 1, 1, 1, 1, 1) &
$\mathcal{U}$(0.5, 1.5) \\
\hline
& \textbf{Shear calibration} \\
$m^{i}$ & (0, 0, 0, 0, 0, 0) & $\mathcal{N}$(0, 0.01) \\ 
\hline
& \textbf{Noise} \\
$10^{10}N^{ij}_{\text{v}\kappa}$ & - & $\mathcal{U}(-100, 100)$
\enddata
\end{deluxetable}

\subsection{Fitting Method}
We employ the $\chi^2$ statistic method to fit the mock data of void-lensing cross-power spectrum, which is given by
\begin{equation}
    \chi^2 = (\bm{D} - \bm{M}(\bm{p}))\textbf{Cov}^{-1}(\bm{D} - \bm{M}(\bm{p})),
\end{equation}
where $\bm{D}$ denotes the data vector of the void-lensing power spectrum, $\bm{M}(\bm{p})$ represents the corresponding theoretical model prediction with parameters $\bm{p}$, and $\textbf{Cov}^{-1}$ is the precision matrix estimated from Equation~(\ref{cov-1}). The likelihood function can then be calculated by $\mathcal{L}$ $\sim$ exp$(-\chi^2/2)$. 

\begin{figure*}[ht]
    \centering
    \includegraphics[width=0.8\linewidth]{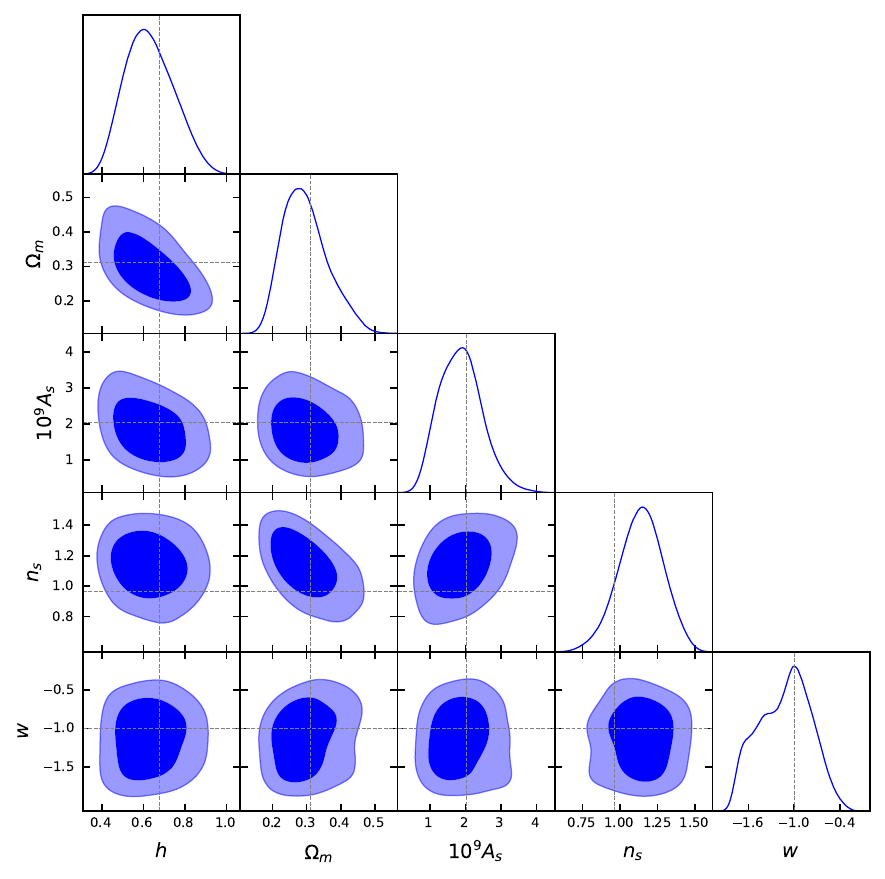}
    \caption{Constraints on the five cosmological parameters derived from the CSST void-lensing analysis over a 100 deg$^2$ survey area. The shaded regions represent the 1$\sigma$ (68.3\%) and 2$\sigma$ (95.5\%) confidence levels. The vertical and horizontal dotted lines indicate the fiducial values of these parameters.}
    \label{fig:cosmopars}
\end{figure*}

We utilize \texttt{emcee}\footnote{\url{https://github.com/dfm/emcee}} \citep{emcee}, a robust and well-tested Python implementation of the affine-invariant ensemble sampler for the MCMC proposed by \cite{MCMC}, to perform the parameter constraints. 
The free parameters, along with their fiducial values and prior assumptions, are summarized in Table~\ref{tab:fit}. The free cosmological parameters include the  reduced Hubble constant $h$, total matter density parameter $\Omega_{m}$, amplitude of the initial power spectrum $A_{s}$, spectral index $n_{s}$ and dark energy equation of state $w$. Since both void and weak-lensing observables are insensitive to the baryon density parameter $\Omega_{b}$, we fix it to the value adopted in the JT1G simulation. 

In addition, we include a set of void parameters, i.e. the void bias $b_{\text{v}}^{i}$ and the void profile parameters $\alpha^{i}$, $\beta^{i}$, $\gamma^{i}$, and $\delta_{\text{cen}}^{i}$ as free parameters for each redshift bin. The multiplicative factors $m^{i}$, photo-$z$ uncertainties $\Delta z^{i}$ and $\sigma_z^{i}$, and noise term $N_{\text{v}\kappa}^{ij}$ are also considered in our fitting process. We adopt flat priors for cosmological parameters, void bias, void profile parameters, and photo-$z$ uncertainties. The Gaussian priors are imposed on the shear calibration parameters, assuming that the CSST photometric survey can offer high imaging quality and excellent control over systematic uncertainties \citep[see e.g.][]{gong19,gong25}.

\subsection{Constraint Results}
In Figure~\ref{fig:cosmopars}, we present the marginalized two-dimensional contours and one-dimensional posterior distributions (PDFs) of the five cosmological parameters constrained from the void-lensing analysis, with the 68.3\% and 95.5\% confidence levels (CLs). The vertical and horizontal dashed lines indicate the fiducial values of the parameters. 

We find that the cosmological parameters can be robustly constrained using the mock data within a 100 deg$^2$ survey area, and the fitting results of the cosmological parameters are consistent with the fiducial values within the 1$\sigma$ CL. For the key parameters $h$, $\Omega_{m}$ and $w$, we obtain constraint accuracies of 18\%, 22\% and 33\%, respectively, corresponding to $h=0.63^{+0.10}_{-0.13}$, $\Omega_{m}=0.296^{+0.049}_{-0.079}$ and $w=-1.12^{+0.39}_{-0.34}$. The constraint results of these cosmological parameters are comparable (e.g. $\Omega_{m}$, $w$) or even tighter (e.g. $h$) than those from the weak lensing only analysis presented in \cite{xiong}, which gives approximately 23\% for $\Omega_{m}$, 30\% for $w$ and 22\% for $h$ using the same mock galaxy catalog. This indicates that the void-lensing cross-correlation encodes important and complementary information about the LSS relative to the weak lensing only measurement, and can provide effective constraints on the relevant cosmological parameters.

These results also demonstrate that our two-dimensional void identification method can reliably construct the 2D void catalog, and the theoretical model can accurately extract cosmological information from the void-lensing correlation. We note that the current constraint accuracies derived from the mock data are limited by the sky area of the light-cone (100 deg$^2$). For the full CSST photometric survey with 17,500 deg$^2$ coverage, the constraints are expected to improve by about an order of magnitude, resulting in accuracies of a few percent-level accuracy. 

\begin{figure}
    \centering
    \includegraphics[width=1\linewidth]{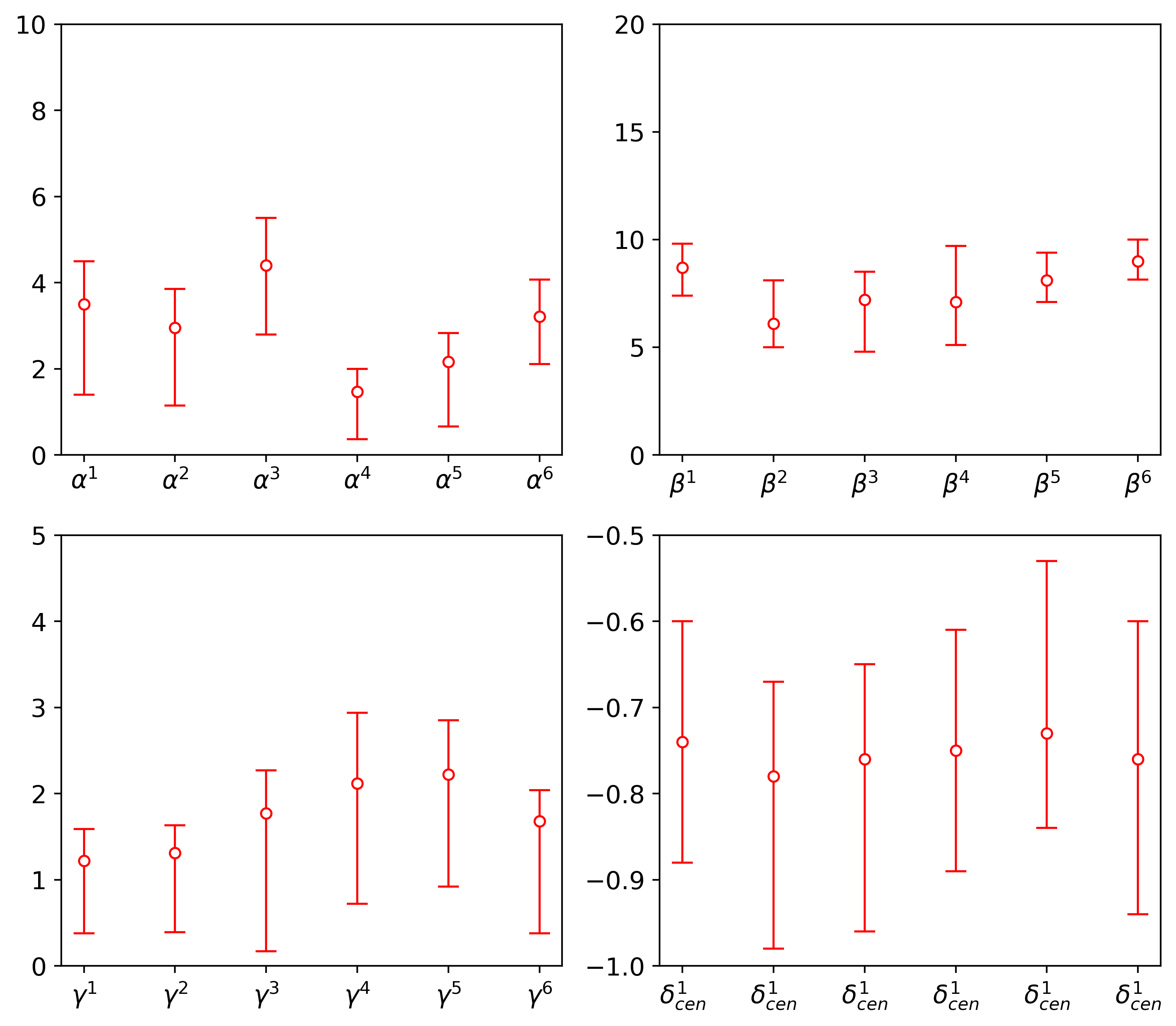}
    \caption{The best-fitting values and corresponding 1$\sigma$ constraints of the parameters $\alpha^{i}$, $\beta^{i}$, $\gamma^{i}$, and $\delta_{\text{cen}}^{i}$ derived from the void lensing analysis.}
    \label{fig:void_p1}
\end{figure}

In Figure~\ref{fig:void_p1}, we show the best-fitting values and corresponding 1$\sigma$ uncertainties for the parameters $\alpha^i$, $\beta^i$, $\gamma^i$, and $\delta^{i}_{\mathrm{cen}}$. Overall, the parameters are well constrained, with most exhibiting relatively small uncertainties compared to their best-fitting amplitudes. For the $\alpha^i$ and $\beta^i$ parameters, the best-fitting values are clustered around $1-5$ and $6-10$, respectively, indicating stable and consistent constraints across the six bins. The $\gamma^i$ parameters exhibit comparatively larger fractional uncertainties, suggesting weaker sensitivity or stronger degeneracy with other parameters in the model. These results are also in good agreement with the previous study using similar galaxy catalog \citep{song5}. The $\delta_{\text{cen}}^{i}$ parameters, ranging from about $-0.8$ to $-0.6$, are consistent with expectations from our 2D void identification procedure. 

\begin{figure}
    \centering
    \includegraphics[width=1\linewidth]{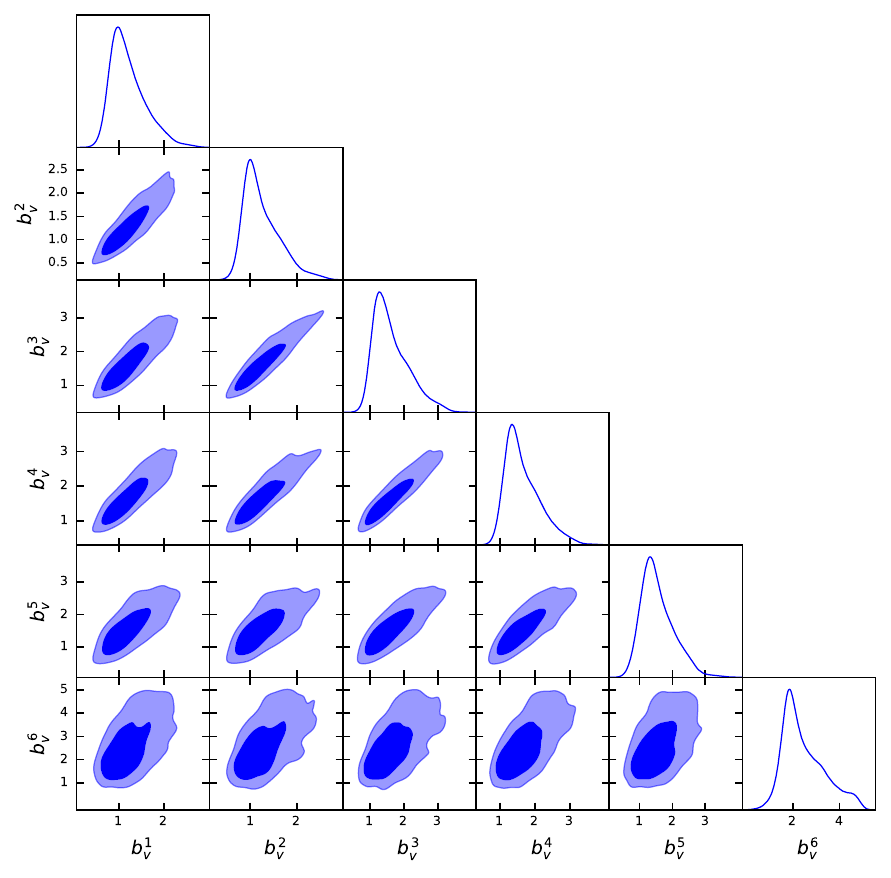}
    \caption{The contour maps (68.3\% and 95.5\% CLs) and 1D PDFs of the void bias for the six tomographic bins from the CSST void-lensing analysis over a 100 deg$^2$ survey area.}
    \label{fig:void_bias}
\end{figure}

In Figure~\ref{fig:void_bias}, we present the constraints on the void bias parameters from the void-lensing analysis for the six tomographic bins, which are shown as contour maps and 1D PDFs. We find that the best-fitting values and errors of the six tomographic bins are $b_{\text{v}}^{1} = 1.21^{+0.23}_{-0.47}$, $b_{\text{v}}^{2} = 1.25^{+0.22}_{-0.47}$, $b_{\text{v}}^{3} = 1.63^{+0.30}_{-0.65}$, $b_{\text{v}}^{4} = 1.65^{+0.30}_{-0.61}$, $b_{\text{v}}^{5} = 1.55^{+0.33}_{-0.60}$, and $b_{\text{v}}^{6} = 2.46^{+0.56}_{-1.1}$, which show an increasing trend as redshift increases.

\begin{figure*}[t]
    \centering
    \includegraphics[width=0.95\linewidth]{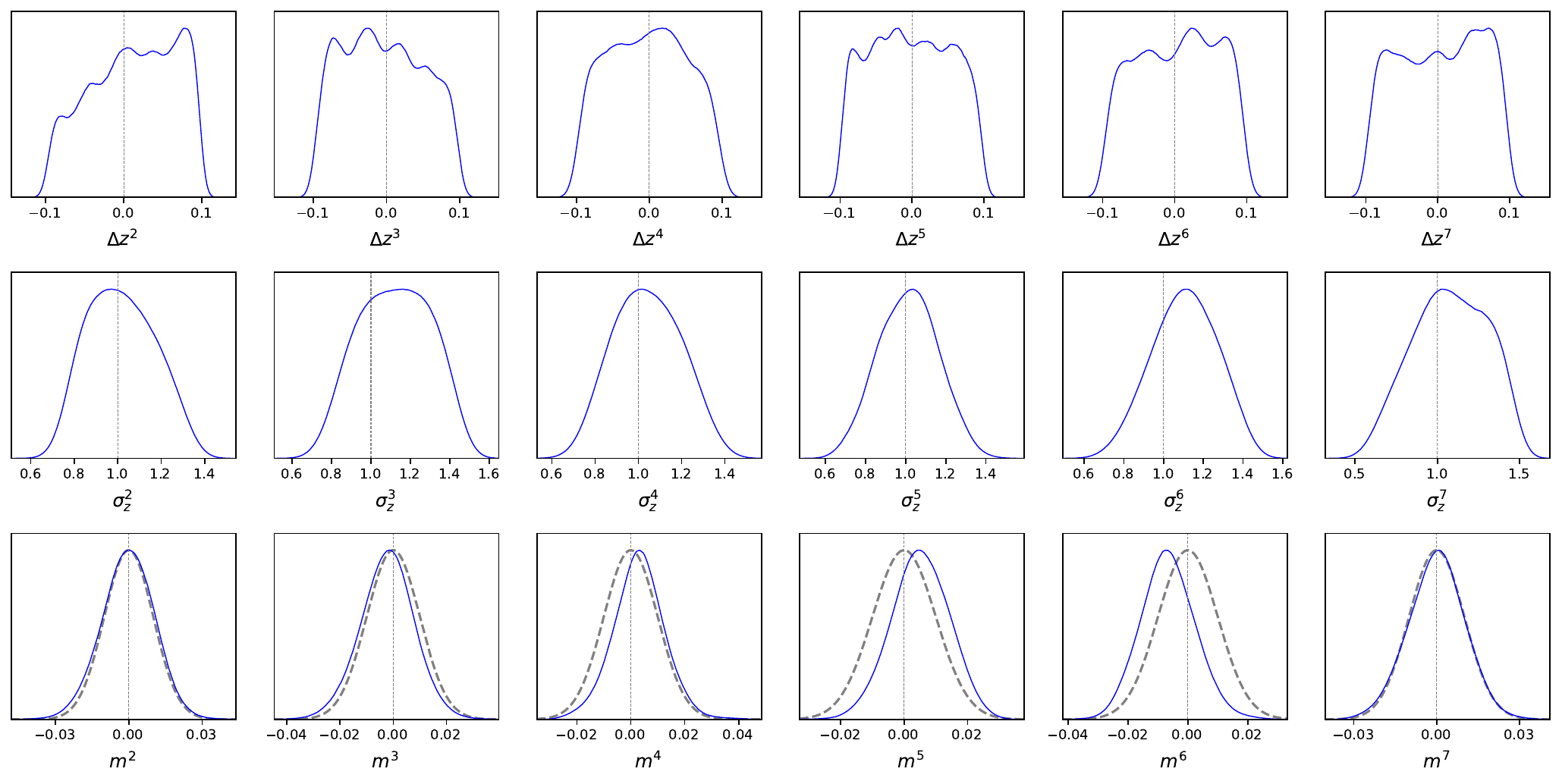}
    \caption{The 1D PDFs of the systematic parameters $\Delta z^{i}$, $\sigma_{z}^{i}$ and $m^{i}$ for the CSST void-lensing constraints in a 100 deg$^2$ survey area for the six tomographic bins. The vertical dotted lines mark the fiducial values of these parameters and the gray dashed lines represent the Gaussian priors.}
    \label{fig:zm}
\end{figure*}

The 1D PDFs of the redshift shift bias $\Delta z^{i}$, stretch factor $\sigma_{z}^{i}$ and multiplicative error $m^i$ are shown in Figure \ref{fig:zm}. The vertical dotted lines mark the fiducial values of these parameters, while the gray dashed curves represent the Gaussian priors of $m^i$. Among these systematic parameters, the stretch factor is relatively well constrained, whereas the constraints on the other two systematic parameters are relatively weak and prior dominated. This can be attributed to the limited survey area and the relatively low statistical power of the dataset used in this work. As demonstrated in previous theoretical studies, the constraint precision on these systematic parameters is expected to improve significantly with the full CSST survey \citep{gong19,Miao,lin,LinAxion}. For the noise term, we obtain consistent constraint results, with $\text{log}_{10}(N_{\text{v}\kappa}^{ij}) \simeq -10$ across all redshift bin combinations $(i,j)$.

\section{Summary} \label{summary}
In this work, we investigate the cosmological constraints from the void-lensing cross-correlation assuming the $w$CDM model for the CSST photometric survey. We construct a mock galaxy catalog covering a sky area of 100 deg$^2$ based on the Jiutian simulations, incorporating the instrumental characteristics and survey strategy of the CSST photometric survey. We divide the galaxy sample into seven photo-$z$ tomographic bins and identify 2D voids in the angular distribution of galaxies within each bin using the Voronoi tessellation and watershed algorithm. We then measure the angular cross-power spectrum between the void distribution and the weak lensing convergence field. The covariance matrix is estimated using the jackknife resampling method combined with the pseudo-$C_{\ell}$ approach to account for the partial sky correction. For the theoretical modeling, we employ the HVDM to model the void-matter cross-power spectrum and adopt the HSW to describe the void density profile. 

After obtaining the mock data, the MCMC technique is employed in the fitting process. We find that the best-fitting values of the cosmological parameters are consistent with the fiducial values within 1$\sigma$ CL. The constraint strength of some cosmological parameters is comparable to or even better than that obtained from the weak lensing only analysis \citep{xiong}, which used the same mock galaxy catalog. These results demonstrate that the void-lensing serves as an effective cosmological probe and a valuable complement to galaxy photometric surveys, particularly for the forthcoming Stage-IV surveys like CSST.

\begin{acknowledgments}
Q.X. and Y.G. acknowledge the support from the CAS Project for Young Scientists in Basic Research (No. YSBR-092), and National Key R\&D Program of China grant Nos. 2022YFF0503404 and 2020SKA0110402. X.L.C. acknowledges the support of the National Natural Science Foundation of China through grant Nos. 11473044 and 11973047 and the Chinese Academy of Science grants ZDKYYQ20200008, QYZDJ- SSW-SLH017, XDB 23040100, and XDA15020200. Q.G. acknowledges the support from the National Natural Science Foundation of China (NSFC No. 12033008). The Jiutian simulations were conducted under the support of the science research grants from the China Manned Space Project with grant No. CMS- CSST-2021-A03. This work is also supported by science research grants from the China Manned Space Project with grant Nos. CMS-CSST-2025-A02, CMS-CSST-2021-B01, and CMS-CSST-2021-A01.
\end{acknowledgments}

\bibliography{sample7}{}
\bibliographystyle{aasjournalv7}



\end{document}